# Who Watches the Watchmen? An Appraisal of Benchmarks for Multiple Sequence Alignment

Stefano Iantorno[1,2,*], Kevin Gori[3,*], Nick Goldman[3], Manuel Gil[4], Christophe Dessimoz[3,†]

[1]Wellcome Trust Sanger Institute, Hinxton, Cambridge, CB10 1SA, UK; [2]U.S. National Institute of Health, National Institute of Allergy and Infectious Diseases, Bethesda, MD 20852, USA; [3]EMBL-EBI European Bioinformatics Institute, Hinxton, Cambridge, CB10 1SA, UK; [4]Center for Integrative Bioinformatics Vienna, Max F. Perutz Laboratories, University of Vienna, Medical University Vienna, 1030 Vienna, Austria

[*]These authors have contributed equally

[†]Corresponding author. E-mail: dessimoz@ebi.ac.uk

**Abstract**

**Multiple sequence alignment (MSA) is a fundamental and ubiquitous technique in bioinformatics used to infer related residues among biological sequences. Thus alignment accuracy is crucial to a vast range of analyses, often in ways difficult to assess in those analyses. To compare the performance of different aligners and help detect systematic errors in alignments, a number of benchmarking strategies have been pursued. Here we present an overview of the main strategies—based on simulation, consistency, protein structure, and phylogeny—and discuss their different advantages and associated risks. We outline a set of desirable characteristics for effective benchmarking, and evaluate each strategy in light of them. We conclude that there is currently no universally applicable means of benchmarking MSA, and that developers and users of alignment tools should base their choice of benchmark depending on the context of application—with a keen awareness of the assumptions underlying each benchmarking strategy.**

# 1. Introduction

Multiple sequence alignment (MSA) has become a common first step in the analysis of sequence data for downstream applications such as comparative genomics, functional analysis and phylogenetic reconstruction. Given their importance, MSA methods need to be objectively validated in order to ensure their output is both accurate and reproducible. Benchmarking is a crucial tool in the assessment of sequence alignment programs, as it allows their developers and users to compare the performance of different aligners objectively, identify strengths and weaknesses and help detect systematic errors in alignments. In recent years, there has been a growing appreciation of the importance of benchmarking measures and datasets to evaluate and critically examine the performance of different MSA software packages, as underscored by a number of recent articles addressing the subject [1-5].

At the same time, and despite these positive developments, the standard approach adopted by the great majority of scientists dealing with sequence alignment has remained reliance on aligners that have long been outperformed in benchmarks [6], or even manual and therefore inevitably subjective intervention in the alignment process [7]. It is unclear whether this is due to the simplicity of use and convenience of long-standing aligners ("historical inertia" [7]), reluctance to move away from customary practice, or unawareness or even distrust of newer, lesser-tested technologies. This trend is particularly worrying in light of the rapid spread of high-throughput technologies and the associated need for automation of analysis pipelines [8]. A reason for this state of affairs might lie the absence of straightforward alignment benchmarking procedure and interpretation. In this chapter, we contribute to overcoming this problem by reviewing present alignment benchmarks, aiming to clarify their strengths and risks for MSA evaluation with a view towards having better (and better-trusted) benchmarks in the future. But before considering benchmarking strategies, we first need to review the alignment objectives we expect them to gauge.

## 1.1. What should sequence aligners strive for?

A conceptual complication lies in the fact that MSAs have multiple and potentially conflicting goals, depending on the biological question of interest [9]. Commonly, the residues aligned are those inferred to be related through homology, i.e. common ancestry. In other contexts, however, the emphasis might be more on functional or structural concordance among residues. A strictly evolutionary interpretation of homology in these cases could be counter-productive, as recognized also by Kemena and Notredame [1], since regions of the protein that carry out the same function or that occupy the same position in the three-dimensional conformation of the protein may have arisen independently by evolutionary convergence. For example, an alignment that pairs structurally analogous, but non-homologous, residues would be informative and therefore "correct" to the structural biologist, although not so to the phylogeneticist. It should however be noted that functional and structural objectives are considerably less precise than the evolutionary objective: while common ancestry is an absolute, binary attribute, similarity in functional or structural role are context-dependent, continuous attributes, thus rendering any reduction to the aligned/unaligned dichotomy subjective at best, ill-defined at worst.

At the same time, the unambiguous nature of the evolutionary objective does not make it automatically easy to pursue (or, as we shall see below, ascertain). Indeed, the evolutionary history of biological sequences is mostly unknown and can only be inferred from present data under the (explicit or implicit) assumption of a model of sequence evolution.

In practice, most MSA methods muddle the distinction among homology-, structure-, or function-motivated alignment by employing strategies anchored in inconsistent objectives. Indeed, almost all well-established aligners assume and exploit evolutionary relationships among the sequences (e.g. by constructing the alignment using an explicitly phylogenetic guide tree and alignment scores derived from models of sequence evolution). Yet many use at the same time structural criteria in their parameters or heuristics, for example by training their parameters using structure-derived reference alignments [10,11]. The complications of the strategies different aligners employ can however be divorced from the measurement of their success, and we wish to make no assumption that an aligner employing one strategy necessarily performs better when assessed according to criteria consistent with its internal methods. In the present context of alignment benchmarking, we therefore treat aligners as "black boxes" and refer the reader interested in the specifics of alignment methods to later chapters.

## 1.2. Aims and desirable properties of alignment benchmarks

As mentioned in the introduction, benchmarks provide ways of evaluating the performance of different MSA packages on standardised input. The output produced by the different programs is compared to the 'correct' solution, the so-called gold standard, that is defined by the benchmark. The extent of similarity between the two then defines the quality of the aligner's performance.

Proper benchmarking is advantageous to both the user and the developer community: the former obtains standardized measures of performance that can be consulted in order to pick the most appropriate MSA tools to address a particular alignment problem, and the latter gains important insight into aspects of the software that need improvement, or new features to be implemented, thus promoting advancement of the field [2].

Which characteristics do benchmarks and the gold standard reference dataset need to satisfy in order to be useful to the user and developer community? Benchmarks can be critically examined by looking at their ability to yield performance measures that reflect the actual biological accuracy (whether defined in terms of shared evolutionary history or structural or functional similarity of the aligned sequence data) of the MSA method. This can most easily be done by defining a set of pre-determined criteria for good benchmarking practice. We follow Aniba *et al*. [2] in their list of desirable properties of benchmarks, which states that a benchmark should be:

- *Relevant*, in that a benchmark should be reflective of actual MSA applications, i.e. tasks carried out by MSA in practice and not in an artificial or hypothetical setting;
- *Solvable*, in that it provides sufficient challenge to differentiate between poor and good performances, while remaining a tractable problem;
- *Scalable*, so that it can grow with the development of MSA programs and sequencing technologies;
- *Accessible*, in order to be widely used by developers and users;
- *Independent* from the methods used by programs under test, as benchmark datasets should avoid any overlap with the heuristics chosen for construction of MSA in order to constitute an objective reference; and

- *Evolving*, to reduce the possibility of developers adapting their programs to a particular test set over time, thus artificially inflating their scores.

Although MSA methods employ different computational solutions to reconstruct sequence alignments, their performance needs to be assessed on the same benchmarks in order to be objectively evaluated and compared. In this chapter, we consider four broad MSA benchmarking strategies (Figure 1):

(i) benchmarks based on simulated evolution of biological sequences, to create examples with known homology;

(ii) benchmarks based on consistency among several alignment techniques;

(iii) benchmarks based on the three-dimensional structure of the proteins encoded by sequence data;

(iv) benchmarks based on knowledge of, or assumption about, the phylogeny of the aligned biological sequences.

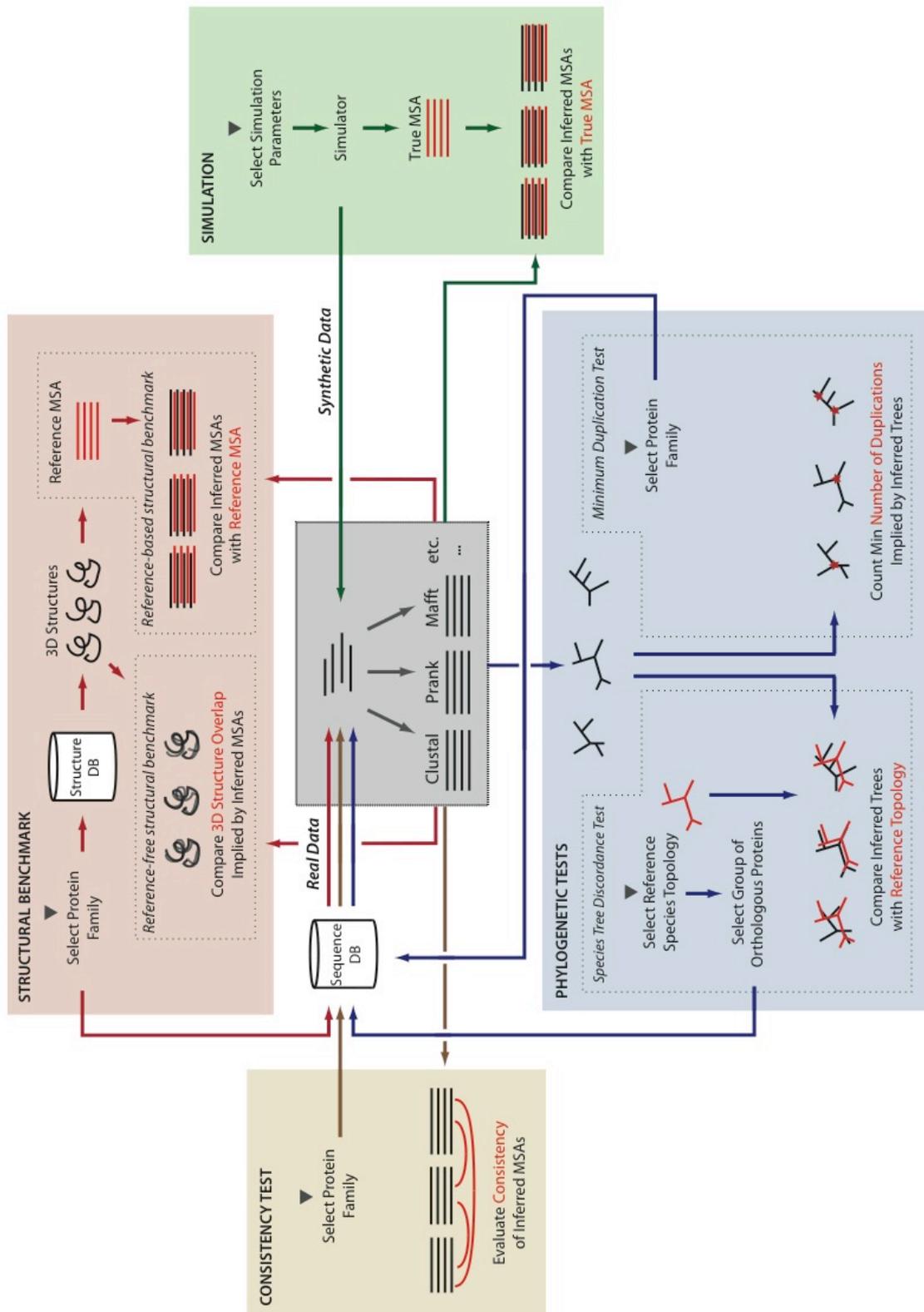

**Figure 1.** *Schematic of the four main MSA benchmarking strategies of this review: for each approach, the benchmarking process starts from the corresponding downward-pointing arrow (▾) and involves alignment by different MSA methods (gray box in centre, illustrating example aligners that may be benchmarked).*

In the remainder of this chapter, we analyse each of these benchmarking approaches to point out their pros and cons, and determine how well they satisfy the criteria defined above and summarised in Table 1.

| Approach | Advantages | Risks | Examples | References |
|---|---|---|---|---|
| **Simulation-based** | · Solvability: 'true' homology is known<br>· Evolving: different scenarios can be modelled<br>· Scalability: new data can be generated *ad libitum* | · Relevance: simulated data might strongly differ from real biological data<br>· Independence: MSA parameters might resemble those used in simulation | Rose<br>DAWG<br>EvolveAGene3<br>iSGv2.0<br>INDELible<br>PhyloSim<br>ALF | [12]<br>[13]<br>[14]<br>[15]<br>[16]<br>[17]<br>[18] |
| **Consistency-based** | · Scalability: not constrained to a particular reference set<br>· Accessibility: tests are easy and quick | · Relevance: consistent MSA methods may be collectively biased<br>· Independence: similar scores might be used in MSA inference | MUMSA<br>HoT | [19,20]<br>[21] |
| **Structure-based** | · Relevance: closely matches a major biological objective of MSA<br>· Independence: empirical data is used as input | · Relevance: limited to structurally conserved regions; biological objective of MSA may vary<br>· Scalability: only applicable to small subset of protein sequences | HOMSTRAD<br>OXBench<br>PREFAB<br>SABMARK<br>BAliBASE 3.0<br>STRIKE | [22,10]<br>[23]<br>[24]<br>[25]<br>[26,11]<br>[27] |
| **Phylogeny-based** | · Relevance: closely matches a major biological objective of MSA<br>· Independence: empirical data is used as input<br>· Scalability: broad array of sequence data can be used as input | · Relevance: biological objective of MSA may vary from phylogenetic reconstruction | Species-tree discordance test<br>Minimum duplication test | [28]<br><br><br>[28] |

**Table 1.** *The advantages and risks of the four approaches to MSA benchmarking. Examples are given of relevant software packages, benchmark databases and tests.*

## 2. Simulated sequences

Given that a major objective of MSA is to identify residues that evolved from a common ancestor, i.e., to optimize for homology in the alignment, one approach to benchmarking involves generating families of artificial sequences by a process of simulated evolution along a known tree. Such simulation-based approaches adopt a probabilistic model of sequence evolution to describe nucleotide substitution, deletion, and insertion rates, while keeping track of 'true' relationships of homology between individual residue positions. Since these are known, a 'true' reference alignment and a test alignment based on the simulated sequence data, assembled by a particular MSA tool of choice, can be compared and measures of accuracy estimated (see below). There are many packages that will perform simulated sequence evolution, including Rose [12], DAWG [13], EvolveAGene3 [14], INDELible [16], PhyloSim [17], REvolver [29] and ALF [18].

To quantify the agreement between the reconstructed alignment and the true alignment (known from the simulation), two measures of accuracy are commonly employed: the sum-of-pairs (SP) and the true column (TC) scores [30]. The former is defined as the fraction of aligned residue pairs that are identical between the reconstructed and true alignment, averaged over all pairwise comparisons between individual sequences; the latter is defined as the fraction of correctly aligned columns that are reproduced in the reconstructed alignment. Given that the TC score considers whole columns in the alignment as comparable units, a single misaligned sequence can reduce the TC score to zero. For this reason, when considering numerous or divergent sequences, the finer-grained SP score is usually used. Yet even the SP score is not without problems. For instance, pairwise comparisons ignore correlations among sequences, meaning that closely related sequences contribute disproportionately more to the SP score than they do to the total phylogenetic information contained in the alignment; this can be misleading in phylogenetic applications. More generally, SP and TC are not proper metrics (they do not satisfy the conditions of symmetry or triangle inequality), which has motivated the recent development of better-founded alternatives [31].

Besides the advantage of knowing the true alignment, the fact that the parameters for simulated sequence evolution are user-defined directly translates into great flexibility to address specific questions or to investigate the effect of individual factors in isolation of others, which is particularly useful to gain insights into the behaviour of complex alignment pipelines. For instance, Löytynoja and Goldman used simulated sequences to expose the systematic underrepresentation of the number of insertions by many aligners, which is especially true as sequence divergence and the number of sequences increases [32].

At the same time, the high level of flexibility afforded by simulation ties in with its biggest drawback: all observations drawn from simulated data depend on the assumptions and simplifications of the model used to generate these data. The vague notion of "realistic simulation" is often used to justify reliance on simulations capturing relevant aspects of real data, but simulations cannot straightforwardly, if at all, account for all evolutionary forces. The risk thus becomes the benchmarking of MSA programs in scenarios of little or no relevance to real biological data. For instance, Golubchik *et al.* investigated the performance of six aligners by simulating sequences in which gaps of constant size were placed in a staggered arrangement across all sequences [33]; although this scenario might be useful to emphasize a more general problem in aligning regions adjacent to gaps, its very artificial nature makes it a poor choice to gauge the extent of that problem on real data.

A further potential risk is the use of simulation settings more favourable to some packages than others [34]. For instance, the selected model of sequence evolution might resemble the underlying model of a particular aligner and thus provide it with an "unfair" advantage (i.e. presumably unrepresentative of typical situations) in the evaluation. Even when the evaluation is conducted in good faith, the high complexity of many MSA aligners—particularly in terms of implicit assumptions and heuristics—can make it challenging to design a fair simulation.

### 3. Consistency among different alignment methods

The key idea behind consistency-based benchmarks is that different good aligners should tend to agree on a common alignment (namely the correct one) whereas poor aligners might make different kinds of mistakes, thus resulting in inconsistent alignments. Confusingly, this notion of consistency among aligners is different from that of consistency-based aligning, which is an alignment strategy that favours MSAs consistent with pairwise alignments [35,36]. In the context of benchmarking, the relevant notion is the former— referred to by Lassmann and Sonnhammer as "inter-consistency", cf. "intra-consistency" for the latter [19].

Practically, benchmarking by consistency among aligners can be implemented using measures such as the overlap score [19], a symmetric variant of sum-of-pairs. From a set of input alignments, all paired aligned residues are determined over all sequences in every alignment. The overlap score for two alignments is calculated by counting the aligned pairs present in both alignments, and dividing by the average number of pairs in the alignments. Hence, two almost identical alignments have an overlap score close to one, while two very different alignments have an overlap score close to zero. Two additional scores based on this concept are the average overlap score, and the multiple overlap score. The average overlap score is simply the mean of the overlap scores measured over all pairs of input alignments, and represents the difficulty of the alignment problem. The multiple overlap score is a weighted sum of all pairs present in a single alignment, with the weight determined by the number of times each pair appears in the whole set of alignments. It is assumed that a high multiple overlap score, gained by an alignment with a high proportion of commonly observed pairs, corresponds to a good performance.

Another score that allows an internal control measure to estimate the consistency of different aligners is the heads-or-tails (HoT) score [21]. This consistency test is based on the assumption that biological sequences do not have a particular direction, and thus that alignments should be unaffected whether the input sequences are given in the original or reversed order. The agreement between the alignments obtained from the original and reversed sequences can be quantified with the overlap measures outlined above.

Both these consistency approaches—consistency among aligners and HoT score—are attractive because they assume no reference alignment or model of sequence of evolution, and thus can be readily and easily employed. Furthermore, high consistency is a necessary quality of a set of accurate aligners, thus making it desirable. The consistency criterion also appeals to the intuitive idea of "independent validation"—although most aligners have many aspects in common and are thus hardly "independent".

The biggest weakness of consistency is that it is no guarantee of correctness: methods can be *consistently wrong*. More subtly, consistency is sensitive to the choice of aligners in the set. This can be partly mitigated by including as many different alignments as possible [19]; nevertheless, it is easy to imagine cases where an accurate alignment, outnumbered by inaccurate, but similar, alignments, will be rated poorly. For instance, a new method solving a problem endemic to existing aligners will have low consistency scores.

Likewise, while low HoT scores can be indicative of considerable alignment uncertainty, the converse is not necessarily true. Hall reported that on simulated data at least, HoT scores tend to overestimate alignment accuracy [37]. That being said, considering the simplicity of HoT's scheme, the correlation he found between HoT and simulation-based measures of alignment accuracy is strikingly high (depending on methods, Pearson ϱ of 87–98%). It remains to be seen whether this will remain the case over time—new aligners might be tempted to exploit HoT's idea in their inference algorithms or parameter optimisation procedures, thus compromising its independence as a benchmarking criterion. For instance, a trivial way of "gaming" the HoT score is to align sequences with "centre-justification" (adding a gap character in the middle of sequences of even-numbered length). Such obviously flawed alignment procedure is nevertheless insensitive to joint sequence reversals, consistently obtaining a perfect HoT score.

## 4. Structural Benchmarks

Benchmarks have also been developed starting from protein structure data. Structural benchmarks are by far the most widely adopted type [2]. Most commonly these employ the superposition of known protein structures as an independent means of alignment, to which alignments derived from sequence analysis can then be compared using the sum-of-pairs and true-column metrics discussed earlier.

Structural benchmarks are naturally highly relevant when sequence alignments are sought to identify structural concordance among amino-acid residues. Yet they are also relevant to an evolutionary interpretation of alignments. Indeed, the biological observation that forms the basis of using structure in the latter context is that homologous proteins often retain structural similarity even when sequence divergence is large [38, Flores, 1993]. Thus, at high levels of divergence, a greater degree of confidence may be placed on alignments based on structural conservation than on sequence similarity. If residues from different proteins can be shown to overlap in three-dimensional space, it is likely (though not certain) that they are homologous. An important advantage of structural benchmarks is that they provide a truly independent, empirically-derived standard to test different alignment algorithms.

A number of structurally-derived benchmark datasets exist. One of the oldest is HOMSTRAD [22,10]. Although not originally intended for benchmarking, this dataset has been extensively used to rate the quality of alignments. The first purpose-built, large-scale structural benchmark was BAliBASE [26,11], which was based on similarity of known protein structures. It is divided into a number of datasets, each suited to test a different alignment problem—for example, greater or lesser sequence diversity, the presence of large insertions or extensions or the presence of repeated elements. Each BAliBASE dataset was constructed by accessing information in structural databases, and alignments were verified by hand, at both the level of individual residues and of overall secondary structure. Other purpose-built structural benchmarks include SABMARK [25] and PREFAB [24], which differ from BAliBASE in that they are derived by automatic means, rather than by manual annotation of protein alignments. Reference sets also exist for RNA structures [39]. For further discussion of these datasets, we direct the reader to reviews by Aniba *et al*. [2], Edgar [3], Kim and Sinha [40], and Thompson *et al*. [4].

Regarding the desirable criterion of independence, although alignment algorithms incorporating structural aspects of sequence data do exist, such as Dynalign [41] and Foldalign [42]—for a more exhaustive discussion of RNA structural alignments, see Gardner *et al*. [39]—the parameters that go into constructing structure-based reference datasets are usually completely detached from the considerations that go into the development of MSA workflows.

Despite the degree of confidence structural alignment confers, it has been observed that sequence alignments used in BAliBASE and PREFAB are not always consistent with known annotations from external sources such as the CATH and SCOP databases, thus calling into question their biological accuracy [3]. Both manual and automated structural benchmark construction face considerable challenges. Manually-curated structural benchmarks, while usually believed to generate more biologically accurate results than automated procedures, might also introduce subjective bias in the alignment. Automated procedures ensure reproducibility, but cannot avoid the existence of debatable parameter choices (e.g. the choice of the minimum spatial distance for two residues to be considered in the same fold) and potential systematic errors.

The non-trivial relationship between structural similarity of residues and alignment highlighted by this study, however, is not the only cause of concern in structural benchmarks. Specifically, structure superpositions used for creating structural benchmarks are often not only based on experimentally derived structures, but also on primary sequence-based procedures such as BLASTP [43] and NORMD [44] which themselves employ amino acid substitution matrices and gap penalty scores, and thus make modelling assumptions about the sequences to be aligned [3]. If these parameters overlap with the parameters employed in MSA methods under test, then reference alignments obtained this way will be biased towards MSA-derived alignments that used those same parameters.

Problems arising from the use in benchmarking of reference alignments derived from structural comparisons can partially be overcome by the direct use of structural measures that are independent of any reference alignment. To evaluate the structure superposition implied by an MSA, Raghava *et al.* [23] adopted scores from a sequence-based multiple structure alignment algorithm [45]. Such structure similarity scores approximate the location of an amino acid in a test alignment by the location of its α-carbon (backbone carbon to which the amino acid side-chain attaches). Two aligned amino acid are then compared by the distance between their chains of α-carbon atoms, estimated by least squares over translations and rotations of their respective 3D protein structures (which are known *a priori*). A simple score is given by the root-mean-square deviations between superposed α-carbon atoms, whereas a more refined score also takes into account the orientation of these atoms [48].

Two final aspects of structural benchmarks further complicate their application in MSA assessment. The fact that reliable annotations exist only for structurally-conserved sequences means that MSA of any region of the genome other than structured protein coding regions—be it intronic, regulatory, natively disordered, or simply poorly annotated—cannot be effectively assessed using existing structural benchmarks [4,40]. This is particularly important given that only a very small fraction of genome sequences encode globular, folded protein domains, and that both structural benchmarks and MSA tools focus mainly on alignment of this very small portion of sequences. The current state of sequencing technologies also means that sequence data come with many artifacts due to sequencing errors, short read length, and/or poor gene prediction models [4,8,46,47] which are only very recently starting to be accounted for in benchmarks [4].

Considering all these complications, it becomes apparent that the map between structure and alignment is neither straightforward nor unequivocal. And indeed, by annotating known domains in reference datasets (or estimating superfamilies when the domain was unavailable), and then comparing annotation agreement in the reference alignments by use of column scores, Edgar found inconsistencies in the assignment of aligned residues to specific secondary structure in both PREFAB and BAliBASE [3].

**5. Phylogenetic tests of alignment**

Our last type of benchmark is phylogenetic tests of alignment. Dessimoz and Gil [28] have recently introduced such tests, developing an MSA assessment pipeline that explicitly takes into consideration phylogenetic relationships within the input sequence data to evaluate the validity of alignment hypotheses generated by different MSA methods.

This approach to benchmarking involves deriving alignments of the test data from different MSA packages as the starting point for tree building. The principle of the tests is simple: the more accurate the resulting tree, the more accurate the underlying alignment is assumed to be. The quality of the tree is measured by its compliance with an auxiliary principle or model; auxiliary in the sense that the additional knowledge introduced be independent of sequence data. So far, two methods have been devised. In the first, referred to as the "species tree discordance test", test alignments are built only from putative orthologous sequences, so that the resulting test trees can be expected to have the same topology as the underlying species tree. Each resulting tree is then compared to a reference species tree, comprising sufficiently divergent species that its branching order is deemed uncontroversial. The best performing aligners are taken to be those that most consistently generate alignments that yield test trees congruent with the species tree. Indeed, it can be expected that averaged over many hundreds or thousands of families, discordance due to non-orthology among the input sequences will affect the performance of all aligners equally, whereas discordance due to alignment error will vary among aligners.

The second method, termed the "minimum duplication test", invokes a parsimony argument to interpret test trees built from alignments of both orthologous and paralogous sequences, favouring trees which require fewer gene duplications to explain the data as more likely to reflect the true evolutionary history of the sequences.

One key advantage of phylogenetic benchmarks is that they provide a way of evaluating gap-rich and variable regions, regions for which structural benchmarks are often not applicable and simulation benchmarks lack realism [28]. In particular, the limited applicability of structural benchmarks to conserved protein core regions has quite possibly caused developers of alignment methods to focus their efforts on improving the performance of their tools on conserved regions at the expense of gap-rich or variable regions. Yet focusing on conserved regions can result in a loss of potentially informative data for multiple sequence alignment [32]. Adopting a simple tree inference method that looks only at presence or absence of gaps as a binary character within a maximum parsimony framework, Dessimoz and Gil reported that gap-only trees are sometimes even more accurate than nucleotide-based trees, thus highlighting the signal lost in neglecting variable or gap-rich regions [28].

At present, phylogeny-based benchmarks are the only ones that can be interpreted to be directly evaluating homology on real data. The premise of this interpretation is that more accurate trees on average necessarily ensue from a higher proportion of homologous positions in alignments on average, and therefore that the former is a good surrogate for the latter. Yet although we view the premise as highly plausible (and indeed fail to see how one could argue the opposite), there is no proof for it. If dismissed altogether, the interpretation has to be weakened so that these phylogeny tests only measure the effect of alignment on phylogenetic inference. In this case, phylogeny-based benchmarks are less meaningful even for other homology-based applications of alignments, such as detecting sites under positive selection [48].

## 6. Conclusions

Benchmarks for MSA applications have arisen in recent years as a crucial tool for bioinformaticians to keep a critical eye on existing software packages and reliably diagnose areas that need further development. The implementation of benchmarks to routinely assess the efficacy and accuracy of MSA methods has clearly provided important insights, and has pointed out to the developer community very serious shortcomings of existing methods that would not otherwise have been so apparent [28,4,49,19]. Each benchmarking solution examined in this chapter—whether simulation-, consistency-, structure-, or phylogeny-based—entails risks of bias and error, but each is also useful in its own right when applied to a relevant problem. It is interesting to note that simulation benchmarks rank MSA methods differently from empirical benchmarks [49,50,32]. It is clear that no single benchmark can be uniformly used to test different MSA methods. Instead, due to both the computational and biological issues raised by the problem of sequence alignment optimization, a multiplicity of scenarios need to be modelled in benchmark datasets.

A telling symptom of the current state of affairs is the fact that subjective manual editing of sequence alignments remains widespread, reflecting perhaps an overall lack of confidence in the performance of automated multiple alignment strategies. The criteria used when editing sequence alignments 'by eye' are vague and may introduce individual biases and aesthetic considerations into sequence alignment [32,9].

In order to ensure reproducibility of experimental results, one of the most important goals of scientific practice, this trend needs to change. Context-dependent, effective benchmarking with well-defined objectives represents a sensible way forward.

**Acknowledgements**

The authors thank Julie Thompson for helpful feedback on the manuscript. CD is supported by SNSF advanced researcher fellowship #136461. This article started as assignment for the graduate course "Reviews in Computational Biology" at the Cambridge Computational Biology Institute, University of Cambridge.